\input harvmac
\noblackbox
 
\input epsf
   
 
\def\journal#1&#2(#3){\unskip, \sl #1\ \bf #2 \rm(19#3) }
\def\andjournal#1&#2(#3){\sl #1~\bf #2 \rm (19#3) }

\def\frac#1#2{{#1\over#2}}

\def\inbar{\,\vrule height1.5ex width.4pt depth0pt}
\def\IC{\relax\hbox{$\inbar\kern-.3em{\rm C}$}}
\def\IR{\relax{\rm I\kern-.18em R}}
\def\IP{\relax{\rm I\kern-.18em P}}
\def\IZ{\relax{\rm I\kern-.18em Z}}
\def\IE{\relax{\rm I\kern-.18em E}}

%
%

%
\catcode`\@=11
\def\slash#1{\mathord{\mathpalette\c@ncel{#1}}}
\overfullrule=0pt

\def\II{{\cal I}}

\def\NN{{\cal N}}
\def\OO{{\cal O}}

\def\ZZ{{\cal Z}}

\def\underrel#1\over#2{\mathrel{\mathop{\kern\z@#1}\limits_{#2}}}

\catcode`\@=12


%

\def \cosh{{\rm cosh}}



\def\unlockat{\catcode`\@=11}
\def\lockat{\catcode`\@=12}

\unlockat


\def\newsec#1{\global\advance\secno by1\message{(\the\secno. #1)}
\global\subsecno=0\global\subsubsecno=0\eqnres@t\noindent
{\bf\the\secno. #1}
\writetoca{{\secsym} {#1}}\par\nobreak\medskip\nobreak}
\global\newcount\subsecno \global\subsecno=0
\def\subsec#1{\global\advance\subsecno
by1\message{(\secsym\the\subsecno. #1)}
\ifnum\lastpenalty>9000\else\bigbreak\fi\global\subsubsecno=0
\noindent{\it\secsym\the\subsecno. #1}
\writetoca{\string\quad {\secsym\the\subsecno.} {#1}}
\par\nobreak\medskip\nobreak}
\global\newcount\subsubsecno \global\subsubsecno=0
\def\subsubsec#1{\global\advance\subsubsecno by1
\message{(\secsym\the\subsecno.\the\subsubsecno. #1)}
\ifnum\lastpenalty>9000\else\bigbreak\fi
\noindent\quad{\secsym\the\subsecno.\the\subsubsecno.}{#1}
\writetoca{\string\qquad{\secsym\the\subsecno.\the\subsubsecno.}{#1}}
\par\nobreak\medskip\nobreak}

\def\subsubseclab#1{\DefWarn#1\xdef
#1{\noexpand\hyperref{}{subsubsection}%
{\secsym\the\subsecno.\the\subsubsecno}%
{\secsym\the\subsecno.\the\subsubsecno}}%
\writedef{#1\leftbracket#1}\wrlabeL{#1=#1}}
\lockat


\newcount\figno
\figno=1
\def\fig#1#2#3{
\par\begingroup\parindent=0pt\leftskip=1cm\rightskip=1cm\parindent=0pt
\baselineskip=11pt
\global\advance\figno by 1
\midinsert
\epsfxsize=#3
\centerline{\epsfbox{#2}}
{\bf Fig.\ \the\figno: } #1\par
\endinsert\endgroup\par
}
\def\figlabel#1{\xdef#1{\the\figno}}
\def\encadremath#1{\vbox{\hrule\hbox{\vrule\kern8pt\vbox{\kern8pt
\hbox{$\displaystyle #1$}\kern8pt}
\kern8pt\vrule}\hrule}}
%
%


\font\cmss=cmss10
\font\cmsss=cmss10 at 7pt
\def\rlx{\relax\leavevmode}
\def\inbar{\vrule height1.5ex width.4pt depth0pt}
\def\IC{\relax\,\hbox{$\inbar\kern-.3em{\rm C}$}}
\def\IN{\relax{\rm I\kern-.18em N}}
\def\IP{\relax{\rm I\kern-.18em P}}
\def\ZZ{\rlx\leavevmode\ifmmode\mathchoice{\hbox{\cmss Z\kern-.4em Z}}
 {\hbox{\cmss Z\kern-.4em Z}}{\lower.9pt\hbox{\cmsss Z\kern-.36em Z}}
 {\lower1.2pt\hbox{\cmsss Z\kern-.36em Z}}\else{\cmss Z\kern-.4em
 Z}\fi}
\def\IZ{\relax\ifmmode\mathchoice
{\hbox{\cmss Z\kern-.4em Z}}{\hbox{\cmss Z\kern-.4em Z}}
{\lower.9pt\hbox{\cmsss Z\kern-.4em Z}}
{\lower1.2pt\hbox{\cmsss Z\kern-.4em Z}}\else{\cmss Z\kern-.4em
Z}\fi}
\def\IZ{\relax\ifmmode\mathchoice
{\hbox{\cmss Z\kern-.4em Z}}{\hbox{\cmss Z\kern-.4em Z}}
{\lower.9pt\hbox{\cmsss Z\kern-.4em Z}}
{\lower1.2pt\hbox{\cmsss Z\kern-.4em Z}}\else{\cmss Z\kern-.4em
Z}\fi}

\def\narrowplus{\kern -.04truein + \kern -.03truein}
\def\narrowminus{- \kern -.04truein}
\def\narrowminussub{\kern -.02truein - \kern -.01truein}

\def\IZ{\relax\ifmmode\mathchoice
{\hbox{\cmss Z\kern-.4em Z}}{\hbox{\cmss Z\kern-.4em Z}}
{\lower.9pt\hbox{\cmsss Z\kern-.4em Z}}
{\lower1.2pt\hbox{\cmsss Z\kern-.4em Z}}\else{\cmss Z\kern-.4em
Z}\fi}
\def\IB{\relax{\rm I\kern-.18em B}}
\def\IC{{\relax\hbox{$\inbar\kern-.3em{\rm C}$}}}
\def\ID{\relax{\rm I\kern-.18em D}}
\def\IE{\relax{\rm I\kern-.18em E}}
\def\IF{\relax{\rm I\kern-.18em F}}
\def\IG{\relax\hbox{$\inbar\kern-.3em{\rm G}$}}
\def\IGa{\relax\hbox{${\rm I}\kern-.18em\Gamma$}}
\def\IH{\relax{\rm I\kern-.18em H}}
\def\II{\relax{\rm I\kern-.18em I}}
\def\IK{\relax{\rm I\kern-.18em K}}
\def\IP{\relax{\rm I\kern-.18em P}}

\font\cmss=cmss10 \font\cmsss=cmss10 at 7pt
\def\IR{\relax{\rm I\kern-.18em R}}


%

%
%
\def\eqnn#1{\xdef #1{(\secsym\the\meqno)}\writedef{#1\leftbracket#1}%
\global\advance\meqno by1\wrlabeL#1}
\def\eqna#1{\xdef #1##1{\hbox{$(\secsym\the\meqno##1)$}}
\writedef{#1\numbersign1\leftbracket#1{\numbersign1}}%
\global\advance\meqno by1\wrlabeL{#1$\{\}$}}
\def\eqn#1#2{\xdef #1{(\secsym\the\meqno)}\writedef{#1\leftbracket#1}%
\global\advance\meqno by1$$#2\eqno#1\eqlabeL#1$$}


\def\boxit#1{\vbox{\hrule\hbox{\vrule\kern8pt
\vbox{\hbox{\kern8pt}\hbox{\vbox{#1}}\hbox{\kern8pt}}
\kern8pt\vrule}\hrule}}
\def\mathboxit#1{\vbox{\hrule\hbox{\vrule\kern5pt\vbox{\kern5pt
\hbox{$\displaystyle #1$}\kern5pt}\kern5pt\vrule}\hrule}}


\lref\EmparanHG{
  R.~Emparan, T.~Harmark, V.~Niarchos and N.~A.~Obers,
  ``Blackfolds in Supergravity and String Theory,''
JHEP {\bf 1108}, 154 (2011).
[arXiv:1106.4428 [hep-th]].
}

\lref\EmparanAT{
  R.~Emparan, T.~Harmark, V.~Niarchos and N.~A.~Obers,
  ``Essentials of Blackfold Dynamics,''
  JHEP {\bf 1003} (2010) 063
  [arXiv:0910.1601 [hep-th]].
}

\lref\GrignaniXM{
  G.~Grignani, T.~Harmark, A.~Marini, N.~A.~Obers and M.~Orselli,
  ``Heating up the BIon,''
JHEP {\bf 1106}, 058 (2011).
[arXiv:1012.1494 [hep-th]].
}

\lref\CallanKZ{
  C.~G.~Callan and J.~M.~Maldacena,
  ``Brane death and dynamics from the Born-Infeld action,''
Nucl.\ Phys.\ B {\bf 513}, 198 (1998).
[hep-th/9708147].
}

\lref\HoweUE{
  P.~S.~Howe, N.~D.~Lambert and P.~C.~West,
  ``The Self-dual string soliton,''
Nucl.\ Phys.\ B {\bf 515}, 203 (1998).
[hep-th/9709014].
}

\lref\LuninMJ{
  O.~Lunin,
  ``Strings ending on branes from supergravity,''
JHEP {\bf 0709}, 093 (2007).
[arXiv:0706.3396 [hep-th]].
}

\lref\NiarchosPN{
  V.~Niarchos and K.~Siampos,
  ``M2-M5 blackfold funnels,''
[arXiv:1205.1535 [hep-th]].
}

\lref\GrignaniMR{
  G.~Grignani, T.~Harmark, A.~Marini, N.~A.~Obers and M.~Orselli,
  ``Thermodynamics of the hot BIon,''
Nucl.\ Phys.\ B {\bf 851}, 462 (2011).
[arXiv:1101.1297 [hep-th]].
}

\lref\EmparanCS{
  R.~Emparan, T.~Harmark, V.~Niarchos and N.~A.~Obers,
  ``World-Volume Effective Theory for Higher-Dimensional Black Holes,''
Phys.\ Rev.\ Lett.\  {\bf 102}, 191301 (2009).
[arXiv:0902.0427 [hep-th]].
}   

\lref\GrignaniIW{
  G.~Grignani, T.~Harmark, A.~Marini, N.~A.~Obers and M.~Orselli,
  ``Thermal string probes in AdS and finite temperature Wilson loops,''
[arXiv:1201.4862 [hep-th]].
}

\lref\rotating{
  V.~Niarchos and K.~Siampos,
  to appear.
}  

\lref\HatefiWZ{
  E.~Hatefi, A.~J.~Nurmagambetov and I.~Y.~Park,
  ``Near-Extremal Black-Branes with $n^3$ Entropy Growth,''
[arXiv:1204.6303 [hep-th]].
}

\lref\KiritsisXC{
  E.~Kiritsis and V.~Niarchos,
  ``Large-N limits of 2d CFTs, Quivers and AdS$_3$ duals,''
JHEP {\bf 1104}, 113 (2011).
[arXiv:1011.5900 [hep-th]].
}

\lref\FreedTG{
  D.~Freed, J.~A.~Harvey, R.~Minasian and G.~W.~Moore,
  ``Gravitational anomaly cancellation for M theory five-branes,''
Adv.\ Theor.\ Math.\ Phys.\  {\bf 2}, 601 (1998).
[hep-th/9803205].
}

\lref\IntriligatorEQ{
  K.~A.~Intriligator,
  ``Anomaly matching and a Hopf-Wess-Zumino term in 6d, N=(2,0) field theories,''
Nucl.\ Phys.\ B {\bf 581}, 257 (2000).
[hep-th/0001205].
}

\lref\BermanEW{
  D.~S.~Berman and J.~A.~Harvey,
  ``The Self-dual string and anomalies in the M5-brane,''
JHEP {\bf 0411}, 015 (2004).
[hep-th/0408198].
}

\lref\GanorVE{
  O.~Ganor and L.~Motl,
  ``Equations of the (2,0) theory and knitted five-branes,''
JHEP {\bf 9805}, 009 (1998).
[hep-th/9803108].
}

\lref\BermanXZ{
  D.~S.~Berman,
  ``Aspects of M-5 brane world volume dynamics,''
Phys.\ Lett.\ B {\bf 572}, 101 (2003).
[hep-th/0307040].
}

\lref\EmparanBR{
  R.~Emparan,
  ``Blackfolds,''
8th chapter of Black Holes in Higher Dimensions  (editor: G. Horowitz), Cambridge University Press.
[arXiv:1106.2021 [hep-th]].
}

\lref\ArmasUF{
  J.~Armas, J.~Camps, T.~Harmark and N.~A.~Obers,
  ``The Young Modulus of Black Strings and the Fine Structure of Blackfolds,''
JHEP {\bf 1202}, 110 (2012).
[arXiv:1110.4835 [hep-th]].
}



\rightline{CCTP-2012-14}
\rightline{CPHT-RR028.0612}
\vskip 1pt
\Title{
}
{\vbox{\centerline{Entropy of the self-dual string soliton}
}}
\medskip
\centerline{Vasilis Niarchos\footnote{$^\flat$}{niarchos@physics.uoc.gr, 
$^\natural$ ksiampos@cpht.polytechnique.fr} and Konstadinos Siampos$^\natural$}
\bigskip
\centerline{{\it $^\flat$Crete Center for Theoretical Physics}}
\centerline{\it Department of Physics, University of Crete, 71303, Greece}
\bigskip
\centerline{{\it $^\natural$Centre de Physique Th\'eorique, \'Ecole Polytechnique}}
\centerline{\it CNRS-UMR 7644, 91128 Palaiseau Cedex, France}
\bigskip\bigskip\bigskip
\centerline{\bf Abstract}
\bigskip

\noindent
We compute the entropy and the corresponding central charge of 
the self-dual string soliton in the supergravity regime using the blackfold
description of the fully localized M2-M5 intersection.

\vfill
\Date{}



\newsec{M2-M5 blackfold funnels}
\seclab\intro

In a recent paper \NiarchosPN\ we provided a new description of the fully localized supergravity 
solution of the M2-M5 intersection in $\IR^{1,10}$
\eqn\introaa{\eqalign{
\vbox{ \offinterlineskip  \halign
{ # & #  & #  &  #  & #  &  # &  #  &  #  & #  & # & # &  #  \cr
      &  0 & 1  & 2   & 3  & 4  & 5   & 6   & 7  & 8 & 9 & 10 \cr 
    \strut &&&&&&&&&&&\cr
M2~:~  & $\bullet$ & $\bullet$ & & & &  & $\bullet$ &&&&    \cr
  \strut &&&&&&&&&&&\cr
M5~:~  & $\bullet$  & $\bullet$ & $\bullet$ & $\bullet$ & $\bullet$ & $\bullet$ & & & & &   \cr}}
      }}
using the blackfold approach \refs{\EmparanCS\EmparanAT-\EmparanHG} (see \EmparanBR\
for a recent review and \refs{\GrignaniXM,\GrignaniIW} for other recent applications in string theory).
In particular, we showed that one can describe the 1/4-BPS intersection \introaa\ in terms of a 
three-funnel (spike) solution of an effective fivebrane worldvolume theory, where a single transverse 
scalar field $z:=x^6$ is activated and behaves as
\eqn\introab{
z(\sigma)=2\pi\frac{N_2}{N_5}\frac{\ell_P^3}{\sigma^2}
~.}
$N_2, N_5$ denote the number of M2 and M5 branes respectively, $\ell_P$ is the Planck scale, and 
$\sigma$ is the radial distance in the directions (2345) transverse to the string intersection but 
parallel to the fivebrane worldvolume. This solution should be contrasted to the self-dual string soliton 
of \HoweUE\ for $N_5=1$, where
\eqn\introac{
z(\sigma)=\frac{2Q_{sd}}{\sigma^2}
~.}
$Q_{sd}$, which is proportional to the number of M2 branes $N_2$, denotes the self-dual 
electric/magnetic charge of the self-dual string. The agreement between the profile \introac\ and
the profile implied by the fully localized supergravity solution was also noticed with alternative
techniques in \LuninMJ.

In this paper we proceed to analyze the thermal version of the spike solution \introab\ following 
(as outlined in \NiarchosPN) the strategy of \GrignaniMR, who considered the analogous situation
for the BIon solution of the D3-F1 system. Since the exact supergravity solution of the black M2-M5 
system is not known, the information we will obtain here is not available currently with any other 
method. 

We are particularly interested in the finite-temperature entropy of the solution from which we can
read off the central charge $c$ of the self-dual string. In the regime, $N_2, N_5\gg 1$, we 
find the following expression 
\eqn\introad{
c\simeq 0.6 \, \frac{N_2^2}{N_5}
~.}
We compare this result with data available from other methods valid in different regimes, and
propose a preliminary interpretation.

\newsec{Thermal spikes}

The blackfold approach describes the (fully localized) supergravity intersection \introaa\ as a spike
solution of an effective fivebrane worldvolume theory. The effective fivebrane worldvolume,
parametrized by coordinates $\hat\sigma^a$ $(a=0,1,\ldots,5)$, is embedded in the ambient eleven 
dimensional flat spacetime whose metric we express as
\eqn\thermaa{
ds_{11}^2=-dt^2+(dx^1)^2+dr^2+r^2 (d\psi^2+\sin^2\psi(d\varphi^2+\sin^2\varphi\, d\omega^2)
+\sum_{i=6}^{10} (dx^i)^2
~.}
The angular coordinates $\psi, \varphi,\omega$ parametrize a round three-sphere. 
In the static gauge
\eqn\thermab{\eqalign{
&t(\hat\sigma^a)=\hat\sigma^0~, ~~ x^1(\hat\sigma^a)=\hat\sigma^1~, ~~ r(\hat\sigma^a)=\hat\sigma^2:=\sigma
\cr
&\psi(\hat\sigma^a)=\hat\sigma^3~, ~~ \varphi(\hat\sigma^a)=\hat\sigma^4~, ~~ \omega(\hat\sigma^a)=\hat\sigma^5~, ~~
x^6(\hat\sigma^a)=z(\sigma)
~.}}
The six dimensional worldvolume is oriented along the directions $(t,x^1,r,\psi,\varphi,\omega)$ and
we are activating only one of the transverse scalars, $x^6=z$. 

It is convenient to define the quantities
\eqn\thermad{
q_2=\frac{16\pi G}{3\Omega_{(3)}\Omega_{(4)}}Q_2~, ~~
q_5=\frac{16\pi G}{3\Omega_{(4)}}Q_5~, ~~
\beta=\frac{3}{4\pi T}~, ~~
\kappa=\frac{q_2}{q_5}
}
where $Q_2, Q_5$ are respectively the charges of the M2 and M5 branes, 
\eqn\thermae{
Q_2=\frac{N_2}{(2\pi)^2\ell_P^3}~, ~~
Q_5=\frac{N_5}{(2\pi)^5 \ell_P^6}
~,}
and $T$ is the temperature. The quantities in \thermad\ are $\sigma$-independent constants that 
parametrize a solution. $\ell_P$ is the eleven dimensional Planck scale, in terms of which 
$16\pi G=(2\pi)^8\ell_P^9$. $\Omega_{(n)}$ denotes the volume of the round $n$-sphere.

For static configurations the equation of motion of the transverse scalar $z(\sigma)$ can be 
determined by extremizing the DBI-like worldvolume action
\eqn\thermaf{
I=\frac{\Omega_{(3)}\Omega_{(4)}L_t L_{x^1}}{16\pi G} \frac{2^{3/2} q_5^3}{\beta^6}\int d\sigma
\sqrt{1+{z'}^2}F(\sigma)
}
where
\eqn\thermag{
F(\sigma)=\sigma^3 \left( \frac{1+\frac{\kappa^2}{\sigma^6}}
{1+\sqrt{1-\frac{4q_5^2}{\beta^6}\left( 1+\frac{\kappa^2}{\sigma^6}\right)}}\right)^{\frac{3}{2}}
\left( -2+\frac{3\beta^6}{2q_5^2} \frac{1+\sqrt{1-\frac{4q_5^2}{\beta^6}
\left(1+\frac{\kappa^2}{\sigma^6}\right)}} {1+\frac{\kappa^2}{\sigma^6}}\right)
~.}
We are using the $+$ branch that has a sensible extremal limit \NiarchosPN. $L_t, L_{x^1}$ denote
the (infinite) length of the $t$, $x^1$ directions. Varying this action with respect to $z(\sigma)$ we 
obtain the equation of motion
\eqn\thermai{
\left( \frac{z'(\sigma) F(\sigma)}{\sqrt{1+z'(\sigma)^2}}\right)'=0
~.}

Ref.\ \NiarchosPN\ noted (in analogy to the BIon case \GrignaniXM) that the reality of the above 
expressions places an upper bound on the temperature $T$
\eqn\thermak{
\beta^3\geq 2 q_5
~.}
Furthermore, it is apparent from \thermag\ that, at any non-zero temperature, a solution will break 
down at 
\eqn\thermal{
\sigma_b=\left( \frac{4 q_2^2}{\beta^6-4 q_5^2} \right)^{\frac{1}{6}}
~.}

This value should be contrasted to the critical value $\sigma_c$ where the validity of the leading 
order approximation breaks down. The description \thermaf\ is the leading term in a 
derivative expansion that assumes the following hierarchy of scales \refs{\GrignaniXM,\NiarchosPN}
\eqn\thermac{
\sigma \gg r_c(\sigma)=q_5^{\frac{1}{3}} \left ( 1+\frac{\kappa^2}{\sigma^6} \right)^{\frac{1}{6}}
~.}
The charge radius $r_c(\sigma)$ is a scale that characterizes the size of the black brane
solution in directions transverse to the effective fivebrane worldvolume. $\sigma_c$ is the critical
value where this hierarchy breaks down, namely $\sigma_c=r_c(\sigma_c)$, or
\eqn\thermam{
\sigma_c=\left( \frac{\pi N_5}{\sqrt 2} \right)^{\frac{1}{3}} \left( 1+\sqrt{1+\frac{64 N_2^2}{N_5^4}} \
\right)^{\frac{1}{6}} \ell_P
~.}
Notice that in contrast to $\sigma_b$ the value of $\sigma_c$ is temperature-independent.

Comparing the characteristic scales $\sigma_b$, $\sigma_c$ we conclude that in the near-extremal 
limit
\eqn\therman{
\frac{\sigma_b}{\sigma_c}=\frac{\sqrt{2} \kappa^{\frac{1}{3}}}{\beta}  
\left( 1+\sqrt{1+\frac{64N_2^2}{N_5^4}}\right)^{-\frac{1}{6}}  \ll 1
}
and the potential breakdown of a thermal spike solution occurs well within the region where
the leading order blackfold approximation cannot be trusted. In that sense, the pathological region
is naturally excised and creates no particular concern. 

{\vbox{{\epsfxsize=100mm \nobreak \centerline{\epsfbox{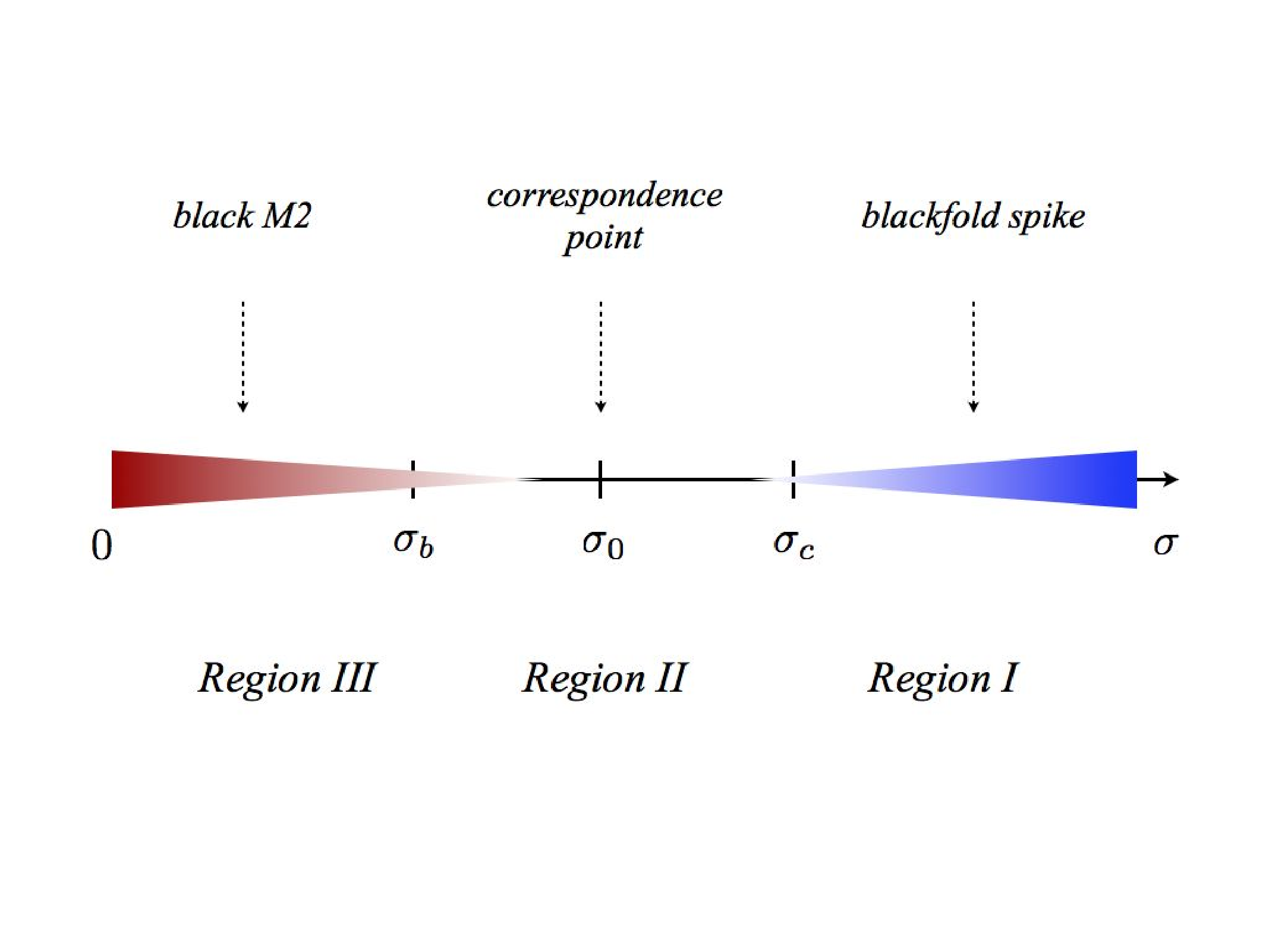}}
\nobreak\vskip-0.8cm {
\it \vbox{ {\bf Figure 1.}
{\it Regions I, II, III on the semi-infinite $\sigma$ line. $\sigma_c$ is the characteristic scale
where the regime of validity of the leading order blackfold approximation ends. $\sigma_b$ is the 
point where a solution of \thermai\ breaks down. $\sigma_0$ denotes the correspondence point 
where the thermodynamic data of a black M2 brane are matched to the thermodynamic data of the 
blackfold spike.
 } }}}}
\bigskip}

The general spike solution of equation \thermai\ is parametrized by a value $\sigma_0$ that 
controls the boundary conditions
\eqn\thermao{
\lim_{\sigma \to+\infty} z(\sigma) =0~, ~~ \lim_{\sigma \to \sigma_0^+}z'(\sigma)=-\infty
~.}
In order to fix the solution uniquely, we need to find the appropriate $\sigma_0$. In the 
extremal case $\sigma_0=0$ and the solution extends all the way to infinity in the $z$ direction.
The addition of temperature modifies the solution and with it $\sigma_0$ which takes a non-zero
temperature-dependent value. In the case where $\sigma_0<\sigma_b$ the solution terminates 
abruptly at the pathological point $\sigma_b$ and the second boundary condition in \thermao\ is
vacuous. For $\sigma_0>\sigma_b$ the solution terminates at 
$\sigma_0$ before reaching $\sigma_b$. From the point of view of the leading order approximation 
the solution is sensible if the termination point is hidden beyong the regime of its validity. In 
that case, the precise value of $\sigma_0$ and the continuation of the solution beyond the
termination point requires data that take us naturally beyond the strict regime of validity of the 
leading order approximation.

In what follows, we propose that the near-extremal spikes are characterized by 
temperature-dependent values $\sigma_0$ that obey the inequalities 
$\sigma_b<\sigma_0<\sigma_c$. The emerging picture
is summarized in Fig.\ 1. As we traverse $\sigma$ from $\sigma=+\infty$ (the location of the 
asymptotic M5 brane) to $\sigma=0$ (the location of the M2 spike) we encounter three different 
regions. In region I, where $\sigma \gg \sigma_c$, the leading order blackfold description 
is valid and the solution is well approximated by a spike solution obeying the boundary conditions
\thermao. At the other extreme, call it region III, where $\sigma\ll \sigma_c$, the notion of a fivebrane 
three-funnel is altogether absent and one should think of the solution more appropriately as a
deformed black M2 brane. In the intermediate region II both descriptions break down and mold 
into each other.

Ref.\ \GrignaniMR\ proposed an approximate scheme for fixing $\sigma_0$, where the 
thermodynamic data of the solutions in regions I and III are matched by extrapolation over the
intermediate region II. The fact that we have to push the blackfold description inside the unreliable 
region II is not unexpected: in order to fix the solution we require data from the emerging M2 spike 
which are only available in this region. 

One obtains two separate gluing conditions in this manner. The first condition comes from the 
matching of the tension of the blackfold solution at $\sigma_0$ to the tension of a black M2 brane. 
The second condition comes from a similar matching of the entropy 
densities. Although a priori independent and derived in a regime where the relevant approximations
break down, the two conditions produce results that agree impressively well with each other.  
We will discuss the issues and viability of this approximate scheme further in section 4. For the 
moment, we assume that this scheme captures sufficiently well the near-extremal thermodynamic
features of the system and proceed to analyze its implications.

\subsec{The spike point of view}

The tension and entropy density of a $\sigma_0$-spike is given 
at any point $\sigma$ by the following formulae \refs{\EmparanHG,\rotating}
\eqn\thermba{
\frac{1}{L_{x^1}}\frac{dM}{dz}=\frac{\Omega_{(3)}\Omega_{(4)}}{16\pi G} \beta^3 \sigma^3 
\frac{F(\sigma)}{F(\sigma_0)}\frac{1+3\cosh^2\alpha}{\cosh^3 \alpha}
~,}
\eqn\thermbb{
\frac{1}{L_{x^1}}\frac{dS}{dz}=\frac{\Omega_{(3)}\Omega_{(4)}}{4 G} \beta^4 \sigma^3 
\frac{F(\sigma)}{F(\sigma_0)}\frac{1}{\cosh^3 \alpha}
~,}
where
\eqn\thermbc{
\cosh \alpha=\frac{\beta^3}{\sqrt 2 q_5}
\sqrt{ \frac{1+\sqrt{1-\frac{4q_5^2}{\beta^6}\left( 1+\frac{\kappa^2}{\sigma^6} \right)}}
{1+\frac{\kappa^2}{\sigma^6}}}
~.}

Expanding these expressions in inverse powers of $\beta$ around the extremal solution 
we obtain 
\eqn\thermbd{
\frac{1}{L_{x^1}}\frac{dM}{dz} \bigg |_{\sigma=\sigma_0^+}=Q_2 
\left ( \sqrt{1+\frac{\sigma_0^6}{\kappa^2}}
+\frac{5 q_2^2}{6\beta^6} \frac{\left( 1+\frac{\sigma_0^6}{\kappa^2}\right)^{\frac{3}{2}}}
{\sigma_0^6}
+\frac{11q_2^4}{8\beta^{12}} \frac{\left( 1+\frac{\sigma_0^6}{\kappa^2}\right)^{\frac{5}{2}}}
{\sigma_0^{12}}
+\OO\left(\beta^{-18}\right) \right )
~,}
\eqn\thermbe{
\frac{1}{L_{x^1}}\frac{dS}{dz} \bigg |_{\sigma=\sigma_0^+}=
\frac{2\pi Q_2}{3} \left ( \frac{2q_2^2\left( 1+\frac{\sigma_0^6}{\kappa^2}\right)^{\frac{3}{2}}}
{\beta^5 \sigma_0^6}
+\frac{3q_2^4 \left( 1+\frac{\sigma_0^6}{\kappa^2} \right)^{\frac{5}{2}}}
{\beta^{11}\sigma_0^{12}}+\OO\left( \beta^{-17} \right)
\right )
~.}

\subsec{The black M2 point of view}

The appropriate description in region III involves a black M2 brane oriented along the directions
(016). The M2 brane has the same charge $Q_2$ and the same temperature $T$ as the M2-M5
system. A perturbative expansion of the tension and entropy density around the extremal limit gives
\eqn\thermca{
\frac{M}{L_{x^1} L_z}=Q_2 \left ( 1+\frac{\sqrt{q_2}}{3\sqrt 2 \beta^3} +\frac{5q_2}{2^6 \beta^6}
+\OO\left(\beta^{-9} \right)
\right )
~,}
\eqn\thermcb{
\frac{S}{L_{x^1} L_z}=\frac{2\pi Q_2}{3} \left ( \frac{\sqrt q_2}{\sqrt 2 \beta^2}
+\frac{3q_2}{16\beta^5}+\OO\left( \beta^{-8} \right)
\right)
~.}

\subsec{Gluing conditions at the correspondence point}

Ref.\ \GrignaniMR\ proposes to match the expressions \thermbd-\thermcb\ order by order in 
the regime where the M2 brane contribution dominates the thermodynamics of the M2-M5
system, namely in the regime where $\kappa^2 \gg \sigma_0^6$. In addition, we require the 
second-order corrections in the expansions \thermbd, \thermbe\ to be subleading. Therefore, in total
\eqn\thermda{
\kappa^2 \gg \sigma_0^6 \gg \frac{q_2^2}{\beta^6}
~.}
These are conditions that restrict our near-extremal analysis to a sufficiently small temperature.

In this regime, the expansions \thermbd, \thermbe\ simplify and the gluing conditions
\eqn\thermdb{
\left( \frac{1}{L_{x^1}}\frac{dM}{dz} \bigg |_{\sigma=\sigma_0^+}\right)_{\rm M2-M5}=
\left( \frac{M}{L_{x^1} L_z}\right)_{\rm M2}
~,}
\eqn\thermdc{
\left( \frac{1}{L_{x^1}}\frac{dS}{dz} \bigg |_{\sigma=\sigma_0^+}\right)_{\rm M2-M5}=
\left( \frac{S}{L_{x^1} L_z} \right)_{\rm M2}
}
imply separately the following temperature expansions of $\sigma_0$
\eqn\thermdd{
\sigma_0^{(M)}=\frac{q_2^{\frac{1}{4}}}{\beta^{\frac{1}{2}}} 
\left( c_1^{(M)} + c_2^{(M)} \frac{q_2^{\frac{1}{2}}}{\beta^3}+\OO\left(\beta^{-6} \right) \right)
, \,
c_1^{(M)}=\left( \frac{25}{2}\right)^{\frac{1}{12}}\simeq 1.234\, ,\,
c_2^{(M)}=-\left( \frac{5^{14}}{2^{79}} \right)^{\frac{1}{12}}\simeq -0.068
,}
\eqn\thermde{
\sigma_0^{(S)}=\frac{q_2^{\frac{1}{4}}}{\beta^{\frac{1}{2}}} 
\left( c_1^{(S)} + c_2^{(S)} \frac{q_2^{\frac{1}{2}}}{\beta^3}+\OO\left(\beta^{-6} \right) \right)
~, ~~ c_1^{(S)}=2^{\frac{1}{4}} \simeq 1.189~,~~ 
c_2^{(S)}=2^{-\frac{17}{4}}\simeq 0.052
~.}
Notice the advertised parametric separation of the scales $\sigma_b \sim \beta^{-1}$ and 
$\sigma_0\sim \beta^{-\frac{1}{2}}$. In addition, 
the coefficients of the leading order terms ($c_1^{(M)}, c_1^{(S)}$) agree well within a 4\% accuracy
despite working in region II that lies beyond the validity of the employed approximations.
There is less agreement in the subleading terms.

The profile of the near-extremal thermal spike $z(\sigma)$ can now be determined 
perturbatively in a small-temperature expansion by solving the equation \thermai\
with boundary conditions \thermao, or equivalently the equation
\eqn\thermdf{
\frac{z'(\sigma) F(\sigma)}{\sqrt{1+z'(\sigma)^2}}=-F(\sigma_0)
~,}
with $\sigma_0=\sigma_0^{(M)}(q_2,\beta)$, or $\sigma_0=\sigma_0^{(S)}(q_2,\beta)$.

\newsec{Entropy and the central charge of the self-dual string soliton}

We are now in position to determine the entropy of the solution. The total entropy is obtained
by integrating the entropy density $\frac{dS}{dz}$ over the full range of $z$ from $z=0$ (in region I)
to $z=+\infty$ (in region III). There are several contributions to this entropy; for instance, contributions 
that scale like $T^5$ from region I where the M5 brane dominates, and contributions that scale like 
$T^2$ from region III where the M2 brane dominates. We are not interested in either of  these
contributions. We are only interested in the contribution of the (1+1) dimensional intersection 
whose leading behavior in the small temperature limit scales like $T$. We propose that this 
contribution can be extracted from the leading order blackfold solution (extrapolated into region II)
by integrating the density \thermbb\ over the range $(\sigma_0,+\infty)$. Namely, it can be
extracted from the integral expression
\eqn\entaa{
\frac{S}{L_{x^1}}=\frac{\Omega_{(3)}\Omega_{(4)} \beta^4}{4G}
\int_{\sigma_0}^{+\infty}d\sigma \, \sigma^3 \frac{F(\sigma)}{\sqrt{F^2(\sigma)-F^2 (\sigma_0)}}
\frac{1}{\cosh^3\alpha}
~.}
The leading order expansion of the integrand in powers of $T$ (with a fixed $\sigma_0$) gives
\eqn\entab{
\frac{S}{L_{x^1}}=\frac{\Omega_{(3)}\Omega_{(4)}}{4G} \frac{q_5^3}{\beta^5}
\int_{\sigma_0}^{+\infty} d\sigma\, \frac{(\kappa^2+\sigma^6)^2}{\sigma^6 \sqrt{\sigma^6-\sigma_0^6}}
+\OO\left(\beta^{-11}\right)
~.}
In this expression we should substitute the leading order behavior of the correspondence point 
\eqn\entac{
\sigma_0=c_1 \frac{q_2^{\frac{1}{4}}}{\beta^{\frac{1}{2}}}
}
where $c_1$ is the numerical coefficient $c_1^{(M)}$ or $c_1^{(S)}$. 

The integral \entab\ is IR divergent in region I $(\sigma \to +\infty)$. The leading contribution from 
that region behaves as $N_5^3 T^5$ and reproduces correctly the entropy of the M5 black brane. We 
are not interested in this contribution, and we can subtract it by introducing an IR cutoff. Alternatively, 
we can perform the integral \entab\ by using hypergeometric functions in which case we remove
the IR divergence by analytic continuation. We have verified that both approaches give the same 
result.

Using the hypergeometric function approach we find
\eqn\entad{
\frac{S}{L_{x^1}}=\frac{\Omega_{(3)}\Omega_{(4)}}{4G} \frac{q_5^3 \sigma_0^4}{\beta^5}
\frac{\Gamma(\frac{1}{3})\Gamma(\frac{1}{6})}{48\sqrt \pi}\,
_2 F_1\left( -2,-\frac{2}{3},-\frac{1}{6};-\frac{\kappa^2}{\sigma_0^6}\right) +\OO\left( \beta^{-11}\right)
~.}
The hypergeometric function is particularly simple in this case
\eqn\entae{
_2 F_1\left( -2,-\frac{2}{3},-\frac{1}{6};-\frac{\kappa^2}{\sigma_0^6}\right)
=1+\frac{8 \kappa^2}{\sigma_0^6}+\frac{8\kappa^4}{5 \sigma_0^{12}}
~.}
Inserting the leading order behavior \entac\ we finally obtain
\eqn\entaf{
\frac{S}{L_{x^1}}=\frac{8\sqrt \pi \Gamma(\frac{1}{3}) \Gamma(\frac{1}{6})} {135 c_1^8}
\frac{N_2^2}{N_5} T+\OO\left( T^4 \right)
~.}

Using the Cardy formula for a two dimensional CFT 
\eqn\entag{
\frac{S}{L_{x^1}}=\frac{\pi c}{6} T
}
we deduce from \entaf\ the following central charge of the self-dual string soliton
\eqn\entai{
c=\frac{16\, \Gamma(\frac{1}{3}) \Gamma(\frac{1}{6})}{45 \sqrt \pi c_1^8} \frac{N_2^2}{N_5}
\simeq 0.6\, \frac{N_2^2}{N_5}
~.}
To obtain the last numerical expression we used $c_1\simeq 1.2$ (see eqs.\ \thermdd, \thermde).

Equation \entai\ is the main result of this paper. In the next and final section we discuss further
the assumptions that went into this computation and compare with other data in the literature 
obtained with different methods.

\newsec{Discussion}

The computation of the entropy of the near-extremal M2-M5 intersection, that led to the result \entai, 
was performed assuming: 
\item{(1)} the validity of an approximation that determines the leading order thermal blackfold spike 
by fixing the boundary conditions \thermao\ at a certain correspondence point $\sigma_0(q_2,T)$, 
obtained by matching data from regions I and III,
\item{(2)} the extrapolation of the descriptions in regions I and III into the region II, where both
are beyond the regime of their validity, and
\item{(3)} that the $\OO(T)$ contribution to the M2-M5 entropy can 
be calculated correctly by integrating the entropy density of the leading order blackfold description 
in regions I and II over the range $(\sigma_0,+\infty)$.

Although these assumptions may appear quite strong, there are preliminary reasons to believe that 
they are sufficiently plausible. First of all, the approximate matching scheme of $(1)$, and the 
extrapolation $(2)$ that it entails, are the natural extension of the successful picture 
in the extremal, 1/4-BPS, case. The extremal case involves a blackfold spike with $\sigma_0=0$.
The leading order blackfold approximation, which again breaks down at the $(N_2,N_5)$-dependent 
point $\sigma_c$ \thermam, reproduces exactly at the tip (at $\sigma=0$) the tension
of the extremal M2 brane \NiarchosPN. This, a priori unexpected, miracle has been observed before
in the DBI treatment of the BIon solution in the case of the D3-F1 system \CallanKZ\ and its blackfold 
treatment in \GrignaniXM. In a companion paper \rotating\ we verify a similar observation 
in the blackfold description of a more complicated extremal rotating M2-M5 intersection with a 
null wave. 

Another encouraging consistency of the assumptions $(1), (2)$ is provided by the non-trivial 
matching of the expansions \thermdd, \thermde. The temperature expansions are the same in 
both cases and the leading order coefficients, $c_1^{(M)}$, $c_1^{(S)}$, agree well within a
4\% discrepancy. This agreement is impressive considering how far from the regime of validity of the 
leading order approximations it was derived. 

As we move away from extremality, it is natural to expect that any deviations between
the exact results and the results obtained with the assumptions $(1), (2)$ will be suppressed by
positive powers of the temperature. For us the crucial question is how such deviations affect the 
order $\OO(T)$ result of the entropy density in equation \entaf. A potential way to probe such 
deviations is to compute the next-to-leading order corrections in the matching scheme $(1)$
by using the next-to-leading order corrected version of the blackfold equations (along the lines
of \ArmasUF).

On dimensional grounds we anticipate that the only way higher order corrections can modify the 
leading order expression \entaf\ is the following. The corrected version of \entaf\ could
take the form of a series expansion
\eqn\discaa{
\frac{S}{L_{x^1}}\sim \frac{1}{G} \frac{q_2^2}{q_5} \left( 1+ a_1 +a_2 +\ldots \right) T+\ldots
}
where the dots at the end of the rhs indicate terms with higher powers of $T$.
Since the IR physics of the system is dominated by the two dimensional CFT at the intersection,
it is natural to expect that the leading temperature dependence of the exact entropy function is 
order $\OO(T)$.

There are three dimensionful parameters that control the blackfold description: $q_2, q_5$ and 
$T$. We can make two independent dimensionless ratios out of these parameters. One of them
involves the temperature, the other is temperature-independent. Accordingly, there are two potential 
corrections carried by the terms $a_i$: corrections that come with positive powers of the temperature 
and corrections that do not depend on the temperature. The former will not contribute to the $\OO(T)$ 
result. The latter are expressed in terms of the dimensionless ratio $\frac{q_2}{q_5^2}$.
Assuming these corrections are suppressed by positive powers of $q_2$, the expansion of the 
$T$-coefficient on the rhs of \discaa\ will be a power series in the ratio 
\eqn\discab{
\frac{1}{\lambda}:=\frac{q_2}{q_5^2}=\frac{4N_2}{N_5^2}
~.}

The intuition behind an expansion in positive powers of $q_2$ (equivalently negative powers of 
$\lambda$) is based on the observation that
$\sigma_0$ is proportional to $q_2^{\frac{1}{4}}$ (see eqs.\ \thermdd, \thermde). Thus, it seems 
sensible to expect that as we decrease $q_2$, the correspondence point $\sigma_0$ comes closer 
to region III and the approximation scheme becomes better. At the same time, we observe that as
we increase $q_2$, $\sigma_c$ also increases pushing region I to the right and 
making the leading order blackfold approximation less accurate. The dependence of $\sigma_c$
on the ratio $\frac{q_2}{q_5^2}$ \thermam\ indeed suggests that the corrections in \discaa\ are of the 
form outlined above. It would be useful and interesting to try to verify explicitly these expectations.

The validity of this picture implies that the leading order treatment of the system in this paper is
relevant in the limit of large $\lambda$, $i.e.$ in the limit $N_2\ll N_5^2$.
We will return to a plausible related interpretation of $\lambda$ in a moment.

Finally, we have assumed in point $(3)$ that the scaling behavior \entaf\ is obtained correctly with an
integral over the regions I and II. This is natural considering the assumptions in points $(1), (2)$. 
The contribution from region III, that involves a black M2 brane, should contribute mostly to the
$\OO(T^2)$ part of the entropy. Unfortunately, we do not have an independent way to verify
this expectation and take it here as part of the assumptions of our overall computational scheme.

\bigskip\noindent
{\it Interpretation of the result and comparison with other data in the literature}
\medskip

The direct computation of the central charge of the self-dual string soliton has been difficult to 
achieve so far due to the lack of a sensible formulation of the non-abelian theory that lives
on the M5 branes. An anomaly computation \refs{\FreedTG\GanorVE\IntriligatorEQ-\BermanEW} 
on the Coulomb branch of the M5 branes suggests that the central charge behaves in the large 
$N_5$ limit as
\eqn\discba{
c\sim N_2 N_5^2~~ {\rm or} ~~ c\sim N_2 N_5
}
depending on the symmetry breaking pattern. The latter is also consistent with a cross section 
scattering calculation \BermanXZ.

From the supergravity point of view the exact localized black M2-M5 intersection has been so far 
inaccessible. A delocalized supergravity solution of the intersection is known \HatefiWZ\ and 
gives a central charge that scales like
\eqn\discbb{
c\sim N_2 N_5
~.}

These results are different from the result in equation \entai.
This is not a contradiction. Both \discba\ and \discbb\ are derived in some abelian limit of the 
theory. In contrast, \entai\ is a fully non-abelian result in the limit $N_2, N_5\gg 1$ 
(with $N_2 \ll N_5^2$).

The next obvious question is the following: does \entai\ have a natural interpretation?
The possibility that the dimensionless ratio \discab\ controls perturbative derivative corrections to our result
suggests that it may be sensible to interpret $\lambda$ as some type of 't Hooft coupling and 
the result \entai\ as the leading order term in a large 't Hooft coupling expansion. Then, it is 
more appropriate to express \entai\ in terms of $N_5$ and $\lambda$, or $N_2$ and $\lambda$. 
In the first case we find
\eqn\discbc{
c \sim 0.04 \, \frac{N_5^3}{\lambda^2}
~.}
This expression exhibits the standard $N_5^3$ scaling of the number of degrees of freedom of the 
M5 brane in the presence of a strong coupling reduction effected by the presence
of the M2 branes. Alternatively, if we express the central charge in terms of $N_2$, $\lambda$
\eqn\discca{
c\sim 0.3 \, \frac{N_2^{\frac{3}{2}}}{\sqrt \lambda}
}
which reproduces the anticipated $N_2^{\frac{3}{2}}$ scaling of the number of degrees of freedom
of the M2 branes (again in the presence of a strong coupling reduction). It is satisfying to see that
this observation recovers sensibly features of both the M2 and M5 branes (with a reduction 
of the degrees of freedom expressed by a negative power of $\lambda$ on both sides). 
 
As a comparative example consider the case of the large-$N$ 't Hooft limit
of the $SU(N)_k$ Wess-Zumino-Witten (WZW) model \KiritsisXC. In that case the natural 't Hooft 
coupling is $\lambda=\frac{N}{k}$ and the central charge scales at leading order in the large 't
Hooft limit as
\eqn\discbd{
c\left(SU(N)_k \right)\sim \frac{N^2}{\lambda}
~.}
Notice that the limit $\lambda \to 0$ is a weak coupling limit in the WZW model. By analogy, and a 
naive use of \discab, the $\lambda \to 0$ limit of the M2-M5 system occurs when $N_5^2 \ll N_2$. 
It seems unlikely that the self-dual string CFT has such a weak coupling limit and we are not 
suggesting this is the case.

The above results make a very specific suggestion for the two dimensional $\NN=(4,4)$ SCFT that 
describes the self-dual string soliton on M5 branes. It would be extremely interesting to find 
independent checks of \entai, to reproduce it from a CFT point of view and to clarify if the above 
interpretation is appropriate or not.

\bigskip
\centerline{\bf Acknowledgements}
\medskip

We would like to thank C.\ Bachas, N.\ Lambert, N.\ Obers and K.\ Sfetsos for useful correspondence and 
discussions. The work of VN was partially supported by the European grants
FP7-REGPOT-2008-1: CreteHEPCosmo-228644, PERG07-GA-2010-268246, and
the EU program ``Thalis'' ESF/NSRF 2007-2013. 
It was also cofinanced by the European Union (European Social Fund, ESF) and Greek national funds
through the Operational Program Education and Lifelong Learning of the National Strategic Reference
Framework (NSRF) under Funding of proposals that have received a positive evaluation in the 3rd and
4th Call of ERC Grant Schemes.
KS has been supported by the ITN programme 
PITN-GA-2009-237920, the ERC Advanced Grant 226371, the IFCPAR CEFIPRA programme 
4104-2 and the ANR programme blanc NT09-573739.

\listrefs
\end